# Rate-Distortion Function for a Heegard-Berger Problem with Two Sources and Degraded Reconstruction sets

Meryem Benammar          Abdellatif Zaidi

### Abstract

In this work, we investigate an instance of the Heegard-Berger problem with two sources and arbitrarily correlated side information sequences at two decoders, in which the reconstruction sets at the decoders are degraded. Specifically, two sources are to be encoded in a manner that one of the two is reproduced losslessly by both decoders, and the other is reproduced to within some prescribed distortion level at one of the two decoders. We establish a single-letter characterization of the rate-distortion function for this model. The investigation of this result in some special cases also sheds light on the utility of joint compression of the two sources. Furthermore, we also generalize our result to the setting in which the source component that is to be recovered by both users is reconstructed in a lossy fashion, under the requirement that all terminals (i.e., the encoder and both decoders) can share an exact copy of the compressed version of this source component, i.e., a common encoder-decoders reconstruction constraint. For this model as well, we establish a single-letter characterization of the associated rate-distortion function.

## I. Introduction

Source coding with decoder side information was first investigated in the seminal paper [1] in which Wyner and Ziv extended the Shannon's standard rate-distortion setup to incorporate decoder side information. Extensions of Wyner and Ziv's result to multiterminal scenarios have motivated extensive investigations, among which Sgarro's work on lossless source coding with side information at many decoders [2] and others. Maybe one of the most important generalizations of Wyner-Ziv setup is the two-user source coding problem of Heegard and Berger [3]. In this setup, a memoryless source $S^n$ has to be reconstructed at two decoders 1 and 2, respectively to within prescribed distortion levels $D_1$ and $D_2$ – Decoder 1 is equipped with side information $Y_1^n$ and Decoder 2 is equipped with side information $Y_2^n$. Heegard and Berger used this setup to model scenarios in which side information may (or may not) be absent. They derived an inner bound on the rate-distortion function $R(D_1, D_2)$ of the model that remains the best bound known to date for the two-user case. Furthermore, this inner bound is optimal for *degraded* side information and for a class of *conditionally less-noisy* side information as shown recently by Timo *et al.* in [4].

Other works proved Heegard and Berger's inner bound to be tight while alleviating the side information ordering constraints. Kimura *et al.* investigated in [5] the *complementary delivery* problem, which consists in coding a two-component source $S^n = (S_1^n, S_2^n)$ such that the component $S_1^n$ can be reproduced at a decoder that knows $S_2^n$ and the component $S_2^n$ can be reproduced at another decoder that knows $S_1^n$. An inner bound on the rate-distortion function was derived in [5] and the rate-distortion function computed in closed form in [6] for some specific settings. Another work for which optimality results are derived without requiring any ordering of side information sequences is the *product of two degraded sources* investigated by Watanabe in [7], where the sources and side information sequences consist in the product of two variables $S = (S_1, S_2)$, $Y_1 = (Y_{1,1}, Y_{1,2})$, and $Y_2 = (Y_{2,1}, Y_{2,2})$ and where $Y_1$ and $Y_2$ are unmatched. The extension of Heegard and Berger's inner bound to more than two users is due to Timo *et al.* in [8] through a successive refinement problem. This inner bound is optimal if the side information sequences are stochastically degraded [9], reversely degraded [10] or, more generally, conditionally less-noisy [4].

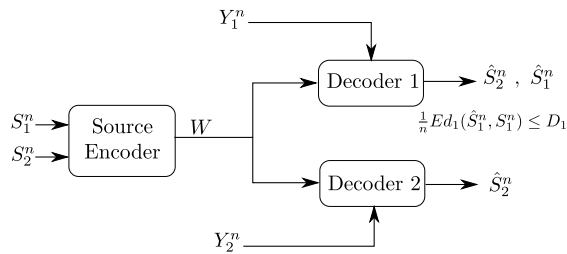

Fig. 1. Heegard-Berger problem with two sources, side information and degraded reconstruction sets.

In this work, we study the source coding model shown in Figure 1. In this model, a memoryless source pair $S^n = (S_1^n, S_2^n)$ is to be encoded and conveyed to two separate decoders. The source component $S_1^n$ has to be reproduced, to within some prescribed average distortion level $D_1$, at only the first decoder, Decoder 1, and the component $S_2^n$ has to be reproduced, at both decoders, losslessly. Decoder 1 observes some side information sequence $Y_1^n$ and Decoder 2 observes side information sequence $Y_2^n$. As opposed to most related works where the side information sequences exhibit certain ordering (e.g., degradedness, reverse degradedness, less-noisiness or the side information being a deterministic function of the sources), in this work we assume that the side information sequences are *arbitrarily correlated* among them, and to the sources.

First, we establish a single-letter characterization of the optimal rate-distortion function $R(D_1)$ of this model. To this end, in particular we derive a converse proof that is tailored specifically for the model with degraded reconstruction sets that we study here. May be somewhat surprisingly, in the optimal coding scheme that we develop for this model, the common description of the sources that is sent to both users involves an auxiliary random variable in addition to the source component $S_2$. The investigation of the role of this variable allows a better understanding of the utility of the *joint* compression of the two sources, as opposed to the *successive* compression that would be expected intuitively to be optimal for this model.

Next, we generalize our results to the setting, shown in Figure 2, in which the source component $S_2^n$ too is to be recovered at both decoders in a lossy fashion, say to within some prescribed average distortion level $D_2$, under the requirement that all terminals (i.e, the encoder and both decoders) can share an exact copy of the compressed version of this source component, i.e., a common encoder-decoders reconstruction constraint of $S_2^n$. For this model as well, we characterize the full rate-distortion function $R(D_1, D_2)$. In the optimal coding scheme that we develop for this model, the compressed version $\hat{S}_2^n$ of the source $S_2^n$, which is sent as a common description, is estimated at both users without utilizing the available side information sequence.

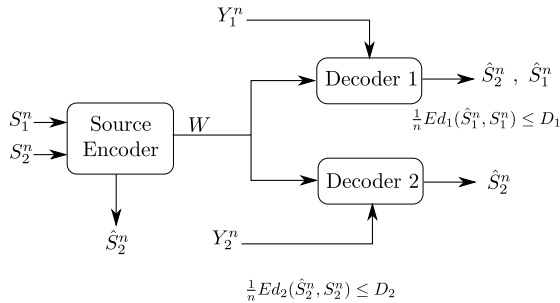

Fig. 2. Heegard-Berger problem with two sources, side information and degraded reconstruction sets.

The model of Figure 2 may be useful for applications in which the source component $S_2^n$ represents some critical information, such as sensitive medical information, and both the sender and the receivers need to share a common compressed version of it, similar to in [11] for the one user case.

### A. Outline and Notation

An outline of the remainder of this paper is as follows. Section II contains formal definitions of the Heegard-Berger problem with degraded reconstruction sets that we study in this work, with and without the aforementioned common reconstruction constraint. In Section III we characterize the rate-distortion function of the model in which the source component $S_2^n$ is reproduced losslessly at both decoders. In this section, we also study some insightful specific cases in which the optimal choices of the random variables that are involved in the rate-distortion function can be characterized explicitly. These specific cases help understand better the role of the common description of the sources that is sent to both users, i.e., the utility of the associated auxiliary random variable. In Section IV, we characterize the rate-distortion function of the model in which the source component $S_2^n$ too is reproduced lossily at both decoders and the encoder and decoders are constrained to share a common compressed version of it.

Throughout the paper we use the following notations. Upper case letters are used to denote random variables, e.g., $X$; lower case letters are used to denote realizations of random variables, e.g., $x$; and calligraphic letters designate alphabets, i.e., $\mathcal{X}$. The probability distribution of a random variable $X$ is denoted by $P_X(x)$. Sometimes, for convenience, we write it as $P_X$. We use the notation $\mathbb{E}_X[\cdot]$ to denote the expectation of random variable $X$. A probability distribution of a random variable $Y$ given $X$ is denoted by $P_{Y|X}$. The set of probability distributions defined on an alphabet $\mathcal{X}$ is denoted by $\mathcal{P}(\mathcal{X})$. The cardinality of a set $\mathcal{X}$ is denoted by $|\mathcal{X}|$. For convenience, the length $n$ vector $x^n$ will occasionally be denoted in boldface notation $\mathbf{x}$. For random variables $X$, $Y$ and $Z$, the notation $X \multimap Y \multimap Z$ indicates that $X$, $Y$ and $Z$, in this order, form a Markov Chain, i.e., $P_{XYZ}(x,y,z) = P_Y(y)P_{X|Y}(x|y)P_{Z|Y}(z|y)$. For integers $i \le j$, we define $[i:j] := \{i, i+1, \dots, j\}$.

## II. Problem Setup and Definitions

Let $(\mathcal{S}_1 \times \mathcal{S}_2 \times \mathcal{Y}_1 \times \mathcal{Y}_2, P_{S_1,S_2,Y_1,Y_2})$ be a discrete memoryless four-source with generic variables $S_1$, $S_2$, $Y_1$ and $Y_2$. Also, let $\hat{\mathcal{S}}_1$ and $\hat{\mathcal{S}}_2$ be two reconstruction alphabets and, for $i \in \{1,2\}$, $d_i$ a distortion measure defined as

$$
\begin{aligned}
d_i \;:\; \mathcal{S}_i \times \hat{\mathcal{S}}_i \;&\to\; \mathbb{R}_+ \\
(s_i, \hat{s}_i) \;&\to\; d_i(s_i, \hat{s}_i) \;.
\end{aligned}
\tag{1}
$$

As we already mentioned, we shall study two related but different models. The model shown in Figure 1 and the model shown in Figure 2. In both models, the side information pair $(Y_1^n, Y_2^n)$ is arbitrarily correlated and to the source pair $(S_1^n, S_2^n)$; Decoder 1 observes only the side information sequence $Y_1^n$ and Decoder 2 observes only the side information sequence $Y_2^n$.

For the model of Figure 1, both decoders recover the source component $S_2^n$ losslessly, and Decoder 1 also recovers the source component $S_1^n$ lossily, to within some prescribed distortion level $D_1$. We shall refer to this problem as the "Heegard-Berger problem with degraded reconstruction sets and one distortion". Formal definitions for this problem are as follows.

*Definition 1:* An $(n, M_n, D_1)$ code for the Heegard-Berger problem with degraded reconstruction sets and one distortion consists of:

- A set of messages $\mathcal{W} \triangleq [1 : M_n]$.

- An encoding function $f$ such that:

$$
\begin{aligned}
f \;:\; \mathcal{S}_1^n \times \mathcal{S}_2^n \;&\to\; \mathcal{W} \\
(S_1^n, S_2^n) \;&\to\; W = f(S_1^n, S_2^n) \;.
\end{aligned}
\tag{2}
$$

- Two decoding functions $g_1$ and $g_2$, one at each user:

$$
\begin{aligned}
g_1 \;:\; \mathcal{W} \times \mathcal{Y}_1^n \;&\to\; \hat{\mathcal{S}}_2^n \times \hat{\mathcal{S}}_1^n \\
(W, Y_1^n) \;&\to\; (\hat{S}_{2,1}^n, \hat{S}_1^n) = g_1(W, Y_1^n) \;,
\end{aligned}
\tag{3}
$$

and

$$
\begin{aligned}
g_2 \; : \; \mathcal{W} \times \mathcal{Y}_2^n &\;\rightarrow\; \hat{\mathcal{S}}_2^n \\
(W, Y_2^n) &\;\rightarrow\; \hat{S}_{2,2}^n = g_2(W, Y_2^n) \; .
\end{aligned}
\tag{4}
$$

The expected distortion of this code is given by

$$
\mathbb{E}\left(d_1(S_1^n, \hat{S}_1^n)\right) \triangleq \mathbb{E}\frac{1}{n}\sum_{i=1}^n d_1(S_{1,i}, \hat{S}_{1,i}) \; .
\tag{5}
$$

The probability of error is defined as

$$
P_e^{(n)} \triangleq \mathbb{P}\left(\hat{S}_{2,1}^n \neq S_2^n \text{ or } \hat{S}_{2,2}^n \neq S_2^n\right) \; .
\tag{6}
$$

□

*Definition 2:* A rate $R$ is said to be $D_1$-achievable for the HB problem with degraded reconstruction sets and one distortion if there exists a sequence of codes $(n, M_n, D_1)$ such that:

$$
\limsup_{n\to\infty} P_e^{(n)} \;=\; 0 \; ,
\tag{7}
$$

$$
\limsup_{n\to\infty} \mathbb{E}\left(d_1(S_1^n, \hat{S}_1^n)\right) \;\leq\; D_1 \; ,
\tag{8}
$$

$$
\liminf_{n\to\infty} \log_2(M_n) \;\geq\; R \; .
\tag{9}
$$

The rate-distortion $R(D_1)$ of this problem is defined by

$$
R(D_1) \triangleq \inf\{R : R \text{ is } D_1\text{-achievable}\} \; .
\tag{10}
$$

□

For the model of Figure 2, the source component $S_2^n$ as well needs to be recovered lossily at both decoders, say to within some prescribed distortion level $D_2$, and all terminals are constrained to share a common compressed version of it. We shall refer to this problem as the "Lossy Heegard-Berger problem with degraded reconstruction sets and common reconstruction". The formal definitions for this problem are similar to the above, with a few additional constraints, and are as follows (we use the same notations for the encoding and decoding functions, for simplicity).

*Definition 3:* An $(n, M_n, D_1, D_2)$ code for the lossy Heegard-Berger problem with degraded reconstruction sets and common reconstruction consists of a set of messages $\mathcal{W} \triangleq [1 : M_n]$, an encoding function $f$ as defined by (2), two reconstructions functions $g_1$ and $g_2$, one at each user, as defined by (3) and (4) respectively, and an additional encoder reconstruction function $g_s$ defined by

$$
\begin{aligned}
g_s \; : \; \mathcal{W} &\;\rightarrow\; \hat{\mathcal{S}}_2^n \\
W &\;\rightarrow\; \hat{S}_{2,s}^n = g_s(W).
\end{aligned}
\tag{11}
$$

The expected distortions of this code are given by

$$
\mathbb{E}\left(d_1(S_1^n, \hat{S}_1^n)\right) \triangleq \mathbb{E}\frac{1}{n}\sum_{i=1}^n d_1(S_{1,i}, \hat{S}_{1,i})
\tag{12}
$$

and

$$
\mathbb{E}\left(d_2(S_2^n, \hat{S}_{2,j}^n)\right) \triangleq \mathbb{E}\frac{1}{n}\sum_{i=1}^n d_2(S_{2,i}, \hat{S}_{2,ji}), \quad \text{for} \quad j = 1, 2.
\tag{13}
$$

The probability of error of this code is given by

$$P_e^{(n)} \triangleq \mathbb{P}\left(\hat{S}_{2,1}^n \neq \hat{S}_{2,s}^n \text{ or } \hat{S}_{2,2}^n \neq \hat{S}_{2,s}^n\right). \tag{14}$$

$\square$

*Definition 4:* A rate $R$ is said to be $(D_1, D_2)$-achievable for the lossy HB problem with degraded reconstruction sets and common reconstruction if there exists a sequence of codes $(n, M_n, D_1, D_2)$ such that:

$$\limsup_{n \to \infty} P_e^{(n)} = 0, \tag{15}$$

$$\limsup_{n \to \infty} \mathbb{E}\left(d_1(S_1^n, \hat{S}_1^n)\right) \leq D_1, \tag{16}$$

$$\limsup_{n \to \infty} \mathbb{E}\left(d_2(S_2^n, \hat{S}_{2,s}^n)\right) \leq D_2, \tag{17}$$

$$\liminf_{n \to \infty} \log_2(M_n) \geq R. \tag{18}$$

The rate-distortion $R(D_1, D_2)$ of this problem is defined by

$$R(D_1, D_2) \triangleq \inf\{R : R \text{ is } (D_1, D_2)\text{-achievable}\}. \tag{19}$$

## III. HB Problem with Degraded Reconstruction Sets and One Distortion

In this section, we study the model of Figure 1.

### A. Main Result

We establish a single-letter characterization of the rate-distortion function $R(D_1)$ of the Heegard-Berger model with degraded reconstruction sets and one distortion shown in Figure 1. The following theorem states the result.

*Theorem 1 (Rate-Distortion Function):* The rate-distortion function $R(D_1)$ of the Heegard-Berger model with degraded reconstruction sets and one distortion is given by

$$R(D_1) = \min_{\mathcal{P}} \max\left\{H(S_2|Y_1) + I(U_0 U_1; S_1|S_2 Y_1), H(S_2|Y_2) + I(U_0; S_1|S_2 Y_2) + I(U_1; S_1|U_0 S_2 Y_1)\right\} \tag{20}$$

where the minimization is over of the set $\mathcal{P}$ of joint conditional pmfs $P_{U_0 U_1 | S_1 S_2}$ that satisfy i) and ii), where:

i) $(U_0, U_1) \multimap (S_1, S_2) \multimap (Y_1, Y_2)$ forms a Markov chain,

ii) there exists a function $\Phi$ such that:

$$\Phi : \mathcal{U}_0 \times \mathcal{U}_1 \times \mathcal{S}_2 \times \mathcal{Y}_1 \to \hat{\mathcal{S}}_1$$

$$(U_0, U_1, S_2, Y_1) \to \hat{S}_1 = \Phi(U_0, U_1, S_2, Y_1)$$

and

$$\mathbb{E}(d_1(S_1, \hat{S}_1)) \leq D_1. \tag{21}$$

**Proof:** The proof of Theorem 1 is given in Appendix A.

The result of Theorem 1 specializes easily to the case of lossless recovery of the source $S_1$ at Decoder 1. The result is stated in the following Corollary.

*Corollary 1 (Lossless Source Coding):* The minimum rate for recovering the source component $S_2$ at Decoder2 and the pair $(S_1, S_2)$ at Decoder 1, all losslessly, is given by

$$R = \min_{P_{U_0|S_1 S_2}} \max \left\{ H(S_1 S_2 | Y_1), H(S_1 S_2 | Y_2) + H(S_1 | Y_1 S_2 U_0) - H(S_1 | Y_2 S_2 U_0) \right\} \tag{22}$$

where the minimization is over conditional $P_{U_0|S_1 S_2}$ satisfying that $U_0 \multimap (S_1, S_2) \multimap (Y_1, Y_2)$ is a Markov chain.

The following remarks help better understanding the result of Theorem 1.

*Remark 1:* The two a.r.v.s that are involved in Theorem 1 can be interpreted as follows. The variable $U_1$ can be interpreted as the individual description of $S_1^n$ that decoder 1 should recover, exactly as in the Heegard and Berger scheme of [3]. The auxiliary variable $U_0$, however, plays a different role. Along with $S_2$, it can be interpreted as a *common* description that is used by both decoders. As it will become clearer (see, e.g., Section IV), the introduction of this auxiliary variable $U_0$ is crucial to the rate-distortion function, and intuitive choices of it, such as $U_0 = \emptyset$ or $U_0 = S_2$, can be *strictly* sub-optimal. This is created by the fact that the side information sequences are not required to exhibit any ordering, and thus, the common description that can be decoded by both decoders, which at least should involve $S_2$, can in fact also involve part or all of $S_1$, depending on the relative strength of the known side information. An extreme case is when $Y_1$ is a degraded version of $Y_2$ and say, the reconstruction of $S_1$ needs to be lossless. In this case, it is clear that the common description should be the entire $(S_2, S_1)$. □

*Remark 2:* In accordance with the insights from Remark 1, the rate-distortion function of Theorem 1 can be expressed equivalently as

$$R(D_1) \geq \min_{\mathcal{P}} \max \left\{ H(S_2 | Y_1) + I(U_0; S_1 | S_2 Y_1), \ H(S_2 | Y_2) + I(U_0; S_1 | S_2 Y_2) \right\} + I(U_1; S_1 | U_0 S_2 Y_1) \tag{23}$$

$$= \min_{\mathcal{P}} \max \left\{ I(U_0 S_2; S_1 S_2 | Y_1), \ I(U_0 S_2; S_1 S_2 | Y_2) \right\} + I(U_1; S_1 | U_0 S_2 Y_1) \tag{24}$$

where the minimization is over the set $\mathcal{P}$ of joint conditional pmfs $P_{U_0 U_1 | S_1 S_2}$ satisfying the aforementioned conditions i) and ii) of Theorem 1.

The readers who are well acquainted with Heegard-Berger type coding may find the expression of the rate-distortion $R(D_1)$ in its form (24) more suitable. In fact, there is a high-level connection among (24) and the result of [3, Theorem 2, p. 733]. More specifically, setting $m = 2$, substituting $X = (S_1, S_2)$ and choosing the random variables $U_T$, for $\emptyset \subset T \subseteq \{1, 2\}$, of [3, Theorem 2, p. 733] as $U_{T=\{1,2\}} = (U_0, S_2)$, $U_{T=\{1\}} = U_1$ and $U_{T=\{2\}} = S_2$, one gets the expression on the RHS of (24). Note, however, that, formally, the achievability result of Theorem 1 can not be obtained by readily applying [3, Theorem 2, p. 733]. The main reason appears to be that the definition of the distortion measure of [3] for the model therein makes it not easily formally amendable to include the definition of the distortion measure for the model that we study here as a specific case of it. Also, the converse proof of [3, Theorem 2, p. 733] is shown there only for the degraded side information case. For the model that we study here, the converse proof is more challenging and involves appropriate bounding steps, use of Csiszár-Körner sum-identity, and identification of the a.r.v. The detailed proof is given in Appendix A. □

*Remark 3:* In most of related works on Heegard-Berger type problems, the side information sequences $Y_1$ and $Y_2$ exhibit a certain ordering (degradedness [9], reverse degradedness [10], or conditional less-noisiness [4]). In this work, we do not make any assumption of this kind on the side information sequences themselves, which can then be arbitrarily correlated. However, our assumption on that the reconstruction sets are degraded appears to be a key ingredient for the optimality result of Theorem 1. Although, intuitively there seem to be high level connections among the setups of degraded side information and degraded reconstruction sets, there are important formal differences; and, in fact, the generalization of the result of Theorem 1 to the case of many decoders with arbitrary side information and degraded reconstruction sets for example appears to be more challenging comparatively. □

## B. Discussion

The result of Corollary 1 has connections with Timo *et. al* [4, Corollary 6.1] and Tian *et. al* [10, Theorem 4]. In both works, one of the decoders recovers only part of the source, e.g., the source component $S_2$ here, while the other decoder seeks to reconstruct the entire source. Further elaboration on this connection can be inferred from the examples of Section III-D. In the case of lossy reconstruction of the source component $S_1$, a result that has connection with our Theorem 1 is due to Timo *et. al* in [4]. In this work, Timo *et. al*, investigated a model in which a function of the source (plays the role of $S_2$ here) is to be recovered at user 2 losslessly, while user 1 requires a lossy reconstruction of the entire source (i.e., $S = (S_1, S_2)$ here). Among other results, they derive the optimal rate-distortion function of this model under some assumptions, which we recall here for completeness.

*Theorem 2 ( [4], Theorem 6):* Under conditionally Less-Noisy side information, i.e. $Y_2 \preccurlyeq Y_1|S_2$ and if $S_2$ is better reconstructed from $Y_1$, i.e. $H(S_2|Y_1) \leq H(S_2|Y_2)$, the rate-distortion function is given by:

$$R(D) = H(S_2|Y_2) + \min_{P_{U_1|S_1 S_2}} I(U_1; S_1|S_2 Y_1) , \qquad (25)$$

where:

- $U_1 \circlearrowleft (S_1, S_2) \circlearrowleft (Y_1, Y_2)$ is a Markov chain

- there exists a function $\Phi$ such that $(\hat{S}_1, \hat{S}_2) = \Phi(U_1, S_2, Y_2)$ verifying the constraint $\mathbb{E}d(S_1 S_2, \hat{S}_1 \hat{S}_2) \leq D$ .

Under the conditions of [4, Theorem 6], if Decoder 2 is able to recover losslessly the source component $S_2$, then so will be Decoder 1, since $H(S_2|Y_1) \leq H(S_2|Y_2)$. However, because the distortion measure for the setup of [4, Theorem 6] is defined differently (joint in [4, Theorem 6] vs. individual in Theorem 1), [4, Theorem 6] can not be implied from Theorem 1, even though our result requires no specific ordering of the side information sequences to hold.

## C. Utility of the Random Variable $U_0$

The introduction of the a.r.v. $U_0$ helps us to leverage the possible imbalance between the side information sequences. In particular, as we already mentioned in Remark 1, informally the common description is such that $(U_0, S_2)$ is richer than only $S_2$ and weaker or equal $(S_2, S_1)$, depending on the side information configuration and the imposed distortion. In this section, we illustrate this aspect through an example.

*Example 1:* Let $(X_1, X_2, X_3, X_4)$ be four independent Ber(1/2) sources. Also, $S_1 = (X_1, X_3)$ and $S_2 = (X_2, X_4)$. Decoder 1 observes the side information $Y_1 = (X_1, X_2, X_4)$ and Decoder 2 observes the side information $Y_2 = X_3$. Both decoders want to reconstruct the second source component $S_2$ losslessly, and Decoder 1 also wants to reconstruct losslessly the first component $S_1$ as depicted in Figure 3.

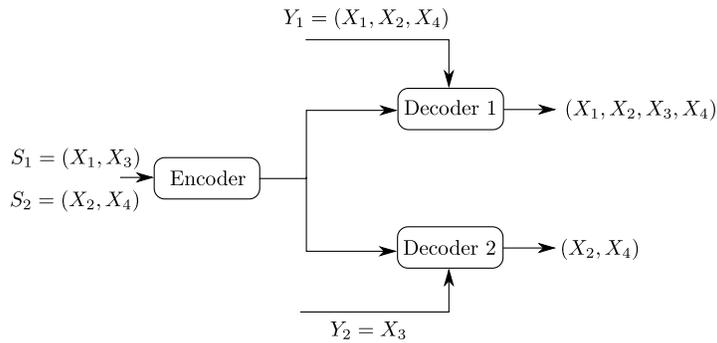

Fig. 3. A binary example of a Heegard-Berger model with degraded reconstruction sets.

For this example, the minimum rate is given by (22) in Corollary 1. It is easy to see that evaluating the right hand side (RHS) of (22) for this example with the choice $U_0 = \emptyset$ yields

$$R_1 = \max\{H(X_1 X_2 X_3 X_4 | Y_1),\ H(X_1 X_2 X_3 X_4 | Y_2)$$

$$+ H(X_1 X_3 | X_2 X_4 Y_1) - H(X_1 X_3 | X_2 X_4 Y_2)\} \tag{26}$$

$$= 3 \text{ bits/sample} \tag{27}$$

Similarly, evaluating the RHS of (22) for this example with the choice $U_0 = S_1 = (X_1, X_3)$ yields

$$R_2 = \max\{H(X_1 X_2 X_3 X_4 | Y_1), H(X_1 X_2 X_3 X_4 | Y_2)\} \tag{28}$$

$$= 3 \text{ bits/sample} \tag{29}$$

However, evaluating the RHS of (22) for this example with the choice $U_0 = X_3$ results in a better (i.e., smaller) rate that is given by

$$R = \max\{H(X_1 X_2 X_3 X_4 | Y_1), H(X_1 X_2 X_4 | X_3) - I(X_1; Y_1 | X_2 X_4 X_3)\} \tag{30}$$

$$= 2 \text{ bits/sample} \tag{31}$$

The rationale behind the choice $U_0 = X_3$, and the reason for which it improves upon the other two choices, are as follows. With the choice $U_0 = X_3$, the source component $X_3$ is conveyed as part of the common description that is intended to both users, although this component is desired only by user 1. This is possible here since user 1 wishes to recover it and user 2 does not lose rate by also recovering it since $Y_2 = X_3$. Conveying $X_3$ as a common description layer is relevant because it saves rate, in comparison to sending it through some individual description layer that is destined to be recovered by only user 1. □

### D. Special Cases

In this section, we evaluate the result of Theorem 1 in some special cases. This evaluation, which requires finding explicit optimal choices of the a.r.v. $U_0$, also allows us to elaborate more on the utility of this auxiliary random variable in these cases. In particular, it will be shown that intuitive choices of $U_0$, such as $U_0 = \emptyset$ or, equivalently, $U_0 = S_2$, can be *strictly* suboptimal even in some of these cases (e.g., the below so-called functional side information case).

*1) Degraded side information case:* The minimum rate, as well as the optimal choices of the a.r.v. $U_0$, depend on the side information ordering. We discuss both cases, degradedness and reverse degradedness.

i) If $Y_2 \multimap Y_1 \multimap (S_1, S_2)$, then the rate-distortion function of Theorem 1 reduces to

$$R_{\text{deg}}(D_1) = \min_{P_{U_1|S_1S_2}} H(S_2|Y_2) + I(U_1; S_1|S_2Y_1) \tag{32}$$

where the minimization is over all $P_{U_1|S_1S_2}$ such that $U_1 \multimap (S_1, S_2) \multimap (Y_1, Y_2)$ and $\mathbb{E}d_1(S_1, \hat{S}_1) \le D_1$ for some $\phi$ such that $\hat{S}_1 = \phi(U_1, Y_1, S_2)$.

A proof of (32) using the result of Theorem 1 is as follows. First, note that, in this degraded case, we have

$$I(U_0S_2; S_1S_2|Y_1) - I(U_0S_2; S_1S_2|Y_2)$$

$$= I(U_0S_2; S_1S_2Y_1) - I(U_0S_2; S_1S_2Y_2) + I(U_0S_2; Y_2) - I(U_0S_2; Y_1) \tag{33a}$$

$$= I(U_0; S_1S_2Y_1) - I(U_0; S_1S_2Y_2) + I(U_0S_2; Y_2) - I(U_0S_2; Y_1) \tag{33b}$$

$$\overset{(a)}{=} I(U_0; S_1S_2) - I(U_0; S_1S_2) + I(U_0S_2; Y_2) - I(U_0S_2; Y_1) \tag{33c}$$

$$= I(U_0S_2; Y_2) - I(U_0S_2; Y_1) \tag{33d}$$

$$\overset{(b)}{\le} 0 \tag{33e}$$

where (*a*) holds since $U_0 \multimap (S_1, S_2) \multimap (Y_1, Y_2)$ is a Markov chain; and (*b*) holds since $Y_2 \multimap Y_1 \multimap (S_1, S_2)$ is a Markov chain. Then, the rate-distortion function of Theorem 1 writes in this case as

$$R_{\text{deg}}(D_1) = \min_{P_{U_0U_1|S_1S_2}} \left\{ \max\{I(U_0S_2; S_1S_2|Y_1), I(U_0S_2; S_1S_2|Y_2)\} + I(U_1; S_1|S_2U_0Y_1) \right\}$$

$$\overset{(c)}{=} \min_{P_{U_0U_1|S_1S_2}} \left\{ I(U_0S_2; S_1S_2|Y_2) + I(U_1; S_1|S_2U_0Y_1) \right\} \tag{34}$$

$$= \min_{P_{U_0U_1|S_1S_2}} \left\{ H(S_2|Y_2) + I(U_0; S_1|S_2Y_2) + I(U_1; S_1|S_2U_0Y_1) \right\} \tag{35}$$

$$= \min_{P_{U_0U_1|S_1S_2}} \left\{ H(S_2|Y_2) + I(U_0; S_1|S_2Y_2) - I(U_0; S_1|S_2Y_1) + I(U_0U_1; S_1|S_2Y_1) \right\} \tag{36}$$

$$\overset{(d)}{=} \min_{P_{U_1|S_1S_2}} \left\{ H(S_2|Y_2) + I(U_1; S_1|S_2Y_1) \right\} \tag{37}$$

where (*c*) holds using (33), and (*b*) holds since using

$$I(U_0; S_1|S_2 Y_2) - I(U_0; S_1|S_2 Y_1) \geq 0 \tag{38}$$

whose proof is similar to (33), we get that

$$H(S_2|Y_2) + I(U_0; S_1|S_2 Y_2) - I(U_0; S_1|S_2 Y_1) + I(U_0 U_1; S_1|S_2 Y_1) \geq H(S_2|Y_2) + I(U_1; S_1|S_2 Y_1) , \tag{39}$$

where the last inequality holds with equality, e.g., with $U_0 = \emptyset$.

The above means that if $Y_2 \multimap Y_1 \multimap (S_1, S_2)$ is a Markov chain then the rate-distortion function of Theorem 1 is attained with the choice $U_0 = \emptyset$, which is then optimal. In this case, the optimal strategy is to use Slepian-Wolf binning [12] to describe the source $S_2$ to both users accounting for $Y_2$ as available side information, and then use Wyner-Ziv binning [1] to describe $S_1$ to user 1 accounting for the pair $(Y_1, S_2)$ as side information. Also, in the specific case of lossless compression, the minimum description rate is given by

$$R_{\text{deg}} = H(S_2|Y_2) + H(S_1|S_2 Y_1) . \tag{40}$$

Finally, we mention that (32) can also be conveyed easily using results from [4, Theorem 6] and [10, Theorem 4].

ii) If $Y_1 \multimap Y_2 \multimap (S_1, S_2)$, then it can be shown relatively easily that the rate-distortion function of Theorem 1 reduces to

$$R_{\text{rev-deg}}(D_1) = \min_{P_{U_1|S_1 S_2}} H(S_2|Y_1) + I(U_1; S_1|S_2 Y_1) \tag{41}$$

where the minimization is over all $P_{U_1|S_1 S_2}$ that satisfy $U_1 \multimap (S_1, S_2) \multimap (Y_1, Y_2)$ and $\mathbb{E} d_1(S_1, \hat{S}_1) \leq D_1$ for some $\phi$ such that $\hat{S}_1 = \phi(U_1, Y_1, S_2)$.

A proof of (41) using the result of Theorem 1 is similar to the aforementioned case. That is,

$$R_{\text{rev-deg}}(D_1) = \min_{P_{U_0 U_1|S_1 S_2}} \left\{ \max\{I(U_0 S_2; S_1 S_2|Y_1), I(U_0 S_2; S_1 S_2|Y_2)\} + I(U_1; S_1|S_2 U_0 Y_1) \right\}$$

$$\stackrel{(a)}{=} \min_{P_{U_0 U_1|S_1 S_2}} \left\{ I(U_0 S_2; S_1 S_2|Y_1) + I(U_1; S_1|S_2 U_0 Y_1) \right\} \tag{42}$$

$$= \min_{P_{U_0 U_1|S_1 S_2}} \left\{ H(S_2|Y_1) + I(U_0 U_1; S_1|S_2 Y_1) \right\} \tag{43}$$

$$= \min_{P_{U_1|S_1 S_2}} \left\{ H(S_2|Y_2) + I(U_1; S_1|S_2 Y_1) \right\} \tag{44}$$

where (*a*) follows since, in this case, we have

$$I(U_0 S_2; S_1 S_2|Y_1) - I(U_0 S_2; S_1 S_2|Y_2) \geq 0, \tag{45}$$

the proof of which is similar to the above.

In the specific case in which the source component $S_1$ as well is recovered losslessly, the minimum rate, which can also be obtained easily using [10, Theorem 4], reduces to

$$R_{\text{rev-deg}} = H(S_1 S_2|Y_1). \tag{46}$$

It is clear that in this case, since the side information $Y_1$ is a degraded version of the side information $Y_2$, if decoder 1 can recover the pair $(S_1, S_2)$, then so can decoder 2. That is, the joint lossless compression of the pair $(S_1, S_2)$ is the rate bottleneck. Hence, the choices $U_0 = \emptyset$ and $U_0 = S_1$ are equally optimal in this case.

Finally, we note that (41) can also be obtained from [10, Theorem 3]. Also, the so-called "side information may be absent" scenarios, i.e., $Y_2 = \emptyset$ or $Y_1 = \emptyset$, are cleary simple special cases of the above i) and ii) respectively. More specifically, if $Y_2 = \emptyset$, the minimum rate is

$$R_{\text{HB}, Y_2 \text{ abs.}} = H(S_2) + H(S_1|Y_1 S_2), \tag{47}$$

and if $Y_1 = \emptyset$, the minimum rate is

$$R_{\text{HB}, Y_1 \text{ abs.}} = H(S_1 S_2). \tag{48}$$

*2) Functional side information:* Another interesting scenario is that of functional side information. We discuss hereafter a few cases for which the rate-distortion function can be computed in a closed form.

i) If $Y_2 = f(S_2)$ and $Y_1$ is *arbitrary,* then the rate-distortion function of Theorem 1 reduces to

$$R(D_1) = \min_{P_{U_1|S_1 S_2}} \max\left\{H(S_2|Y_1),\, H(S_2|Y_2)\right\} + I(U_1; S_1|Y_1 S_2). \tag{49}$$

where the minimization is over all conditionals $P_{U_1|S_1 S_2}$ such that $U_1 \multimap (S_1, S_2) \multimap (Y_1, Y_2)$ and $\mathbb{E} d_1(S_1, \hat{S}_1) \leq D_1$, with $\hat{S}_1 = \phi(U_1, Y_1, S_2)$ for some function $\phi$.

A proof of (49) is as follows. We have

$$\min_{P_{U_0 U_1|S_1 S_2}} \left\{\max\{I(U_0 S_2; S_1 S_2|Y_1), I(U_0 S_2; S_1 S_2|Y_2)\} + I(U_1; S_1|S_2 U_0 Y_1)\right\}$$

$$= \min_{P_{U_0 U_1|S_1 S_2}} \left\{\max\{H(S_2|Y_1) + I(U_0; S_1|S_2 Y_1), H(S_2|Y_2) + I(U_0; S_1|S_2 Y_2)\} + I(U_1; S_1|S_2 U_0 Y_1)\right\} \tag{50}$$

$$= \min_{P_{U_0 U_1|S_1 S_2}} \left\{\max\{H(S_2|Y_1) + I(U_0 U_1; S_1|S_2 Y_1), H(S_2|Y_2) + I(U_0; S_1|S_2 Y_2)\} + I(U_1; S_1|S_2 U_0 Y_1)\}\right\} \tag{51}$$

$$\overset{(a)}{=} \min_{P_{U_0 U_1|S_1 S_2}} \max\left\{H(S_2|Y_1) + I(U_0 U_1; S_1|S_2 Y_1), H(S_2|Y_2) + I(U_0; S_1|S_2) + I(U_1; S_1|S_2 U_0 Y_1)\right\} \tag{52}$$

$$\overset{(b)}{=} \min_{P_{U_1|S_1 S_2}} \left\{\max\{H(S_2|Y_1), H(S_2|Y_1)\} + I(U_1; S_1|S_2 Y_1)\right\} \tag{53}$$

where (a) holds since $Y_2 = f(S_2)$ in this case and (b) holds using the fact that

$$I(U_0 U_1; S_1|S_2 Y_1) \geq I(U_1; S_1|S_2 Y_1) \tag{54}$$

and

$$I(U_0; S_1|S_2) + I(U_1; S_1|S_2 U_0 Y_1) \overset{(c)}{=} I(U_0; S_1 Y_1|S_2) + I(U_1; S_1|S_2 U_0 Y_1) \tag{55}$$

$$\geq I(U_0; S_1|Y_1 S_2) + I(U_1; S_1|S_2 U_0 Y_1) \tag{56}$$

$$= I(U_0 U_1; S_1|S_2 Y_1) \tag{57}$$

$$\geq I(U_1; S_1|S_2 Y_1) \tag{58}$$

where (c) follows using the Markov chain $U_0 \multimap (S_1, S_2) \multimap Y_1$; and noticing that the inequalities are attained with equality with the choice $U_0 = \emptyset$.

The above means that if $Y_2 = f(S_2)$ then the rate-distortion function of Theorem 1 is attained with the choice $U_0 = \emptyset$, which is then optimal. In the specific case in which the source component $S_1$ as well is recovered losslessly, the minimum rate is given by

$$R = \max\Big\{H(S_2S_1|Y_1),\ H(S_2|Y_2) + H(S_1|Y_1S_2)\Big\}. \tag{59}$$

ii) If $Y_1 = f(S_2)$ and $Y_2$ is arbitrary, then the minimum rate of Corollary 1 reduces to

$$R = \max\Big\{H(S_1S_2|Y_1),\ H(S_1S_2|Y_2)\Big\}. \tag{60}$$

A proof of (60) using the result of Corollary 1 is as follows. We have

$$\min_{P_{U_0|S_1S_2}} \Big\{\max\{I(U_0S_2; S_1S_2|Y_1), I(U_0S_2; S_1S_2|Y_2)\} + H(S_1|S_2U_0Y_1)\Big\}$$

$$= \min_{P_{U_0|S_1S_2}} \max\Big\{H(S_1S_2|Y_1), I(U_0S_2; S_1S_2|Y_2) + H(S_1|S_2U_0Y_1)\Big\} \tag{61}$$

$$= \max\Big\{H(S_1S_2|Y_1),\ \min_{P_{U_0|S_1S_2}} \{I(U_0S_2; S_1S_2|Y_2) + H(S_1|S_2U_0Y_1)\}\Big\} \tag{62}$$

$$= \max\Big\{H(S_1S_2|Y_1),\ \min_{P_{U_0|S_1S_2}} \{H(S_1S_2|Y_2) + H(S_1|S_2U_0Y_1) - H(S_1|S_2U_0Y_2)\}\Big\} \tag{63}$$

$$\overset{(a)}{=} \max\Big\{H(S_1S_2|Y_1),\ \min_{P_{U_0|S_1S_2}} \{H(S_1S_2|Y_2) + H(S_1|S_2U_0) - H(S_1|S_2U_0Y_2)\}\Big\} \tag{64}$$

$$= \max\Big\{H(S_1S_2|Y_1),\ \min_{P_{U_0|S_1S_2}} \{H(S_1S_2|Y_2) + I(S_1; Y_2|S_2U_0)\}\Big\} \tag{65}$$

$$\overset{(b)}{=} \max\Big\{H(S_1S_2|Y_1), H(S_1S_2|Y_2)\Big\} \tag{66}$$

where (a) follows since $Y_1 = f(S_2)$, and (b) follows since

$$I(S_1; Y_2|S_2U_0) \geq 0\ , \tag{67}$$

with the inequality holding with equality for the choice $U_0 = S_1$.

The above means that an optimal choice of the minimizing $U_0$ in the minimum description rate of Corollary 1 is $U_0 = S_1$. It is important to observe that, by opposition to the choice $U_0 = S_1$, the choice $U_0 = \emptyset$ therein yields a rate that is generally strictly suboptimal, as it satisfies

$$R(U_0 = \emptyset) = \max\{H(S_1S_2|Y_1),\ H(S_2|Y_2) + H(S_1|Y_1S_2)\} \tag{68}$$

$$\geq \max\{H(S_1S_2|Y_1),\ H(S_1S_2|Y_2)\}. \tag{69}$$

An instance of the the so-called complementary delivery problem corresponds to the specific case in which $Y_2 = S_1$ and $Y_1 = S_2$ (see also [5]). From the above, it can be easily seen that the minimum rate in this case is given by

$$R_{\text{comp-deliv}} = \max\{H(S_2|S_1),\ H(S_1|S_2)\}\ . \tag{70}$$

Here as well, the choice $U_0 = \emptyset$ in the minimum description rate of Corollary 1 is generally strictly sub-optimal as it yields

$$R_{\text{comp-deliv}}(U_0 = \emptyset) = H(S_2|S_1) + H(S_1|S_2) \tag{71}$$

$$\geq \max\{H(S_2|S_1), \, H(S_1|S_2)\} \tag{72}$$

$$= R_{\text{comp-deliv}}. \tag{73}$$

For example, if $S_1$ and $S_2$ are both Ber(1/2) sources and are independent, we have $R_{\text{comp-deliv}}(U_0 = \emptyset) = 2$ bits/sample and $R_{\text{comp-deliv}} = 1$ bit/sample.

## IV. HB Problem with Degraded Reconstruction Sets and Common Reconstruction

In this section, we generalize our results to the model of Figure 2 in which, as we already mentioned, the source component $S_2^n$ as well is now required to be recovered lossily at both deocders, to within some average prescribed distortion level $D_2$, under the requirement that all terminals (i.e., the encoder and both decoders) can share an exact copy of the compressed version of it. The main result of this section, a full characterization of the rate-distortion function of the model of Figure 2, is stated in Section IV-B. In section IV-A, we briefly review and comment on some related results on the role of side information for binning and/or estimation – the reader may find this helpful for a better understanding of the result of Section IV-B.

### A. Role of Side Information, Binning and/or Estimation

In source coding problems with side information at the decoder, side information may be utilized for binning and/or estimation, depending on the configuration. For example, in the standard Wyner-Ziv setup [1] with source $S$ and arbitrarily correlated side information $Y$ available non-causally only at the decoder, the side information is utilized both for binning and for the estimation of the reconstruction. This is reflected through the associated rate-distortion function, which is given by

$$R_{\text{WZ}} = \min_{P_{V|S Y}} I(V; S|Y) \tag{74}$$

where $P_{VSY}$ is such that:

1) $\quad V \multimap S \multimap Y \tag{75}$

2) $\quad \exists \quad \hat{S} \triangleq \phi(V; Y) \quad \text{s.t} \quad \mathbb{E}d(S, \hat{S}) \leq D. \tag{76}$

The side information plays a similar role, but with a generally stronger binning leading to a better rate, if it is also given to the encoder (i.e., the standard conditional rate-distortion problem [13]); and it plays a less important role (only estimation) if it is given only causally to the decoder but not to the encoder [14]. If the encoder is constrained to produce an exact copy of the decoder's reconstruction, referred to as "common encoder-decoder reconstruction constraint" in [11], side information can be used for binning but not for estimation – as otherwise, the encoder, which does not know the side information, can not estimate the decoder's reconstruction. This is reflected through the associated rate-distortion function, which in this case reduces to [11]

$$R_{\text{CR}} = \min_{P_{\hat{S}SY}} I(\hat{S}; S|Y) \tag{77}$$

where $P_{\hat{S}SY}$ is such that:

1) $\hat{S} \multimap S \multimap Y$ $\qquad\qquad\qquad\qquad\qquad\qquad\qquad\qquad (78)$

2) $\mathbb{E}d(S, \hat{S}) \leq D$ $\qquad\qquad\qquad\qquad\qquad\qquad\qquad\qquad (79)$

The role of side information is less easy to understand, comparatively, in extensions of Wyner-Ziv's setup to multi-terminal settings, such as the Heegard-Berger problem with common reconstruction constraint. For example, for the Heegard Berger model with *common receivers reconstruction only* shown in Figure 4, Vellambi and Timo [15] observed that side information can be used at the estimation phase. ( [15] also shows that side information is only useful for binning if the side information pair are degraded and a *common source-receivers reconstruction* constraint is imposed as in Figure 5). The reader may also refer to the related work in [16] where Ahmadi *et. al* characterize the rate distortion function of a Heegard-Berger model with degraded side information in which the encoder is constrained to be able to produce copies of the decoders' reconstructions, without imposing that these decoders' reconstructions be identical to each other with high probability – see Figure 6.

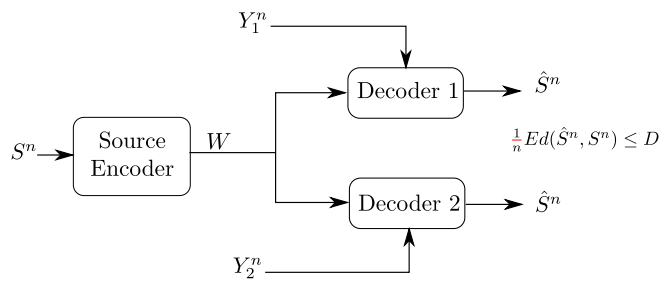

Fig. 4. Heegard-Berger problem with common receiver reconstruction only

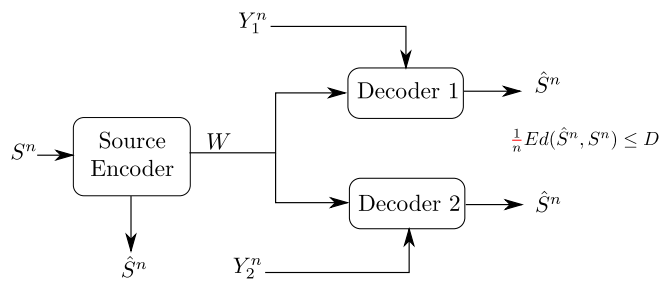

Fig. 5. Heegard-Berger problem with common source receiver reconstruction

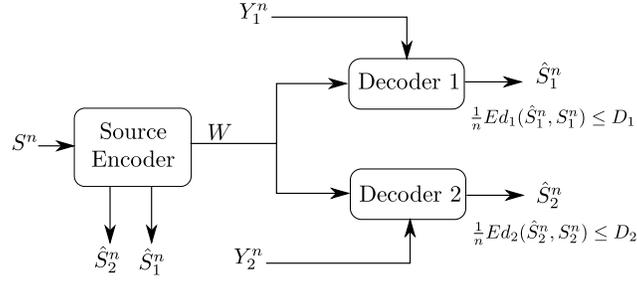

Fig. 6. Heegard-Berger problem with common reconstruction

## B. Rate-Distortion Function

Recall the definitions 3 and 4 of section II. The following theorem characterizes the rate-distortion function of the Heegard-Berger model with degraded reconstruction sets and common reconstruction shown in Figure 2.

*Theorem 3 (The rate-distortion function):* The rate-distortion function $R(D_1, D_2)$ of the Heegard-Berger model with degraded reconstruction sets and common reconstruction shown in Figure 2 is given by

$$R(D_1, D_2) = \min_{\mathcal{P}} \max \left\{ I(U_0 \hat{S}_2; S_1 S_2 | Y_1), \ I(U_0 \hat{S}_2; S_1 S_2 | Y_2) \right\} + I(U_1; S_1 S_2 | Y_1 \hat{S}_2 U_0) \tag{80}$$

where the minimization is over of the set $\mathcal{P}$ of joint conditional pmfs $P_{U_0 U_1 \hat{S}_2 | S_1 S_2}$ that satisfy i), ii), and *iii*) where:

i) $(U_0, U_1, \hat{S}_2) \multimap (S_1, S_2) \multimap (Y_1, Y_2)$ forms a Markov chain,

ii) there exists a function $\Phi$ such that:

$$\Phi \ : \ \mathcal{U}_0 \times \mathcal{U}_1 \times \hat{S}_2 \times \mathcal{Y}_1 \to \hat{S}_1$$

$$(U_0, U_1, \hat{S}_2, Y_1) \to \hat{S}_1 = \Phi(U_0, U_1, \hat{S}_2, Y_1)$$

and

$$\mathbb{E}(d_1(S_1, \hat{S}_1)) \le D_1, \tag{81}$$

iii)

$$\mathbb{E}(d_2(S_2, \hat{S}_2)) \le D_2. \tag{82}$$

**Proof:** The proof of Theorem 3 is given in Appendix B.

In the following remarks, we elaborate more on Theorem 3 and its connection to Theorem 1.

*Remark 4:* The result of Theorem 3 can be seen as a generalization of that of Theorem 1, in the sense that setting $D_2 = 0$ in Theorem 3 one recovers Theorem 1. Also, a characterization of the rate-distortion function for the model of Figure 4 can be readily obtained from the result of Theorem 3 by setting $S_1 = \emptyset$. In this sense, Theorem 3 can also be seen as a generalization of [15, Theorem 1] to the case in which one of the decoders also recovers an individual description. □

*Remark 5:* The coding scheme that we use for the proof of achievability of Theorem 3 is similar to that of Theorem 1, with the main difference being that, for the encoding and decoding of the source component $s_2^n$, the side information sequences $y_1^n$ and $y_2^n$ are used for the binning stage, but not for the estimation stage– However, they are used for both binning and estimation for the encoding/decoding of the source component $s_1^n$.

*Remark 6:* Related to Remark 5, given that in the model of Figure 2 that we study in this section the encoder can produce the desired compression version $\hat{S}_2$ that all terminals want to share, the reader may wonder whether, for the achievability proof, the model with common reconstruction constraint of Figure 2 that we study in this section could be connected to the model with one distortion that we study in Section III by viewing the source pair of Figure 1 as $(S_1, \hat{S}_2)$, instead of $(S_1, S_2)$, and the role of the encoder therein as that of conveying the component $\hat{S}_2$ to both decoders in a lossless fashion and the component $S_1$ to only Decoder 1 in a lossy fashion. Although this high level connection holds, caution should be exercised as the resulting model would still be different from that of Figure 1 in Section III, e.g., in that the encoder would have additional correlated side information $S_2$.

## Acknowledgement

The authors would like to thank Roy Timo for the helpful discussions about the model studied in this paper.

## Appendix A
## Proof of Theorem 1

### A. Proof of converse

In this section, we show that if $R$ is $D_1$-achievable, then $R \geq R(D_1)$.

Assume that $R$ is $D_1$-achievable. Let then $\Phi$ such that $\hat{S}_1^n = \Phi(W, Y_1^n)$ and where $\frac{1}{n} \mathbb{E}(d_1(S_1^n, \hat{S}_1^n)) \leq D_1$. Also, let $\epsilon_n$ be an asymptotically vanishing sequence such that:

$$\forall j \in \{1, 2\} \quad s.t \quad H(S_2^n | W Y_j^n) \leq n\epsilon_n \ . \tag{83}$$

First, note that we have

$$n R \geq I(W; S_1^n S_2^n | Y_1^n) \tag{84}$$

$$= I(W; S_2^n | Y_1^n) + I(W; S_1^n | Y_1^n S_2^n) \tag{85}$$

$$\overset{(a)}{\geq} H(S_2^n | Y_1^n) + I(W; S_1^n | Y_1^n S_2^n) - n\epsilon_n \tag{86}$$

$$\overset{(b)}{=} nH(S_2 | Y_1) + \sum_{i=1}^n I(W; S_{1,i} | Y_{1,i} S_{2,i} Y_1^{i-1} Y_{1,i+1}^n S_2^{i-1} S_{2,i+1}^n S_{1,i+1}^n) - n\epsilon_n \tag{87}$$

$$= nH(S_2 | Y_1) + \sum_{i=1}^n I(W Y_1^{i-1} Y_{1,i+1}^n S_2^{i-1} S_{2,i+1}^n S_{1,i+1}^n; S_{1,i} | Y_{1,i} S_{2,i}) - n\epsilon_n \tag{88}$$

$$\overset{(c)}{=} nH(S_2 | Y_1) + \sum_{i=1}^n I(W Y_1^{i-1} Y_{1,i+1}^n S_2^{i-1} S_{2,i+1}^n S_{1,i+1}^n Y_{2,i+1}^n; S_{1,i} | Y_{1,i} S_{2,i}) - n\epsilon_n \tag{89}$$

$$\geq nH(S_2|Y_1) + \sum_{i=1}^{n} I(WY_1^{i-1}Y_{2,i+1}^n Y_{1,i+1}^n S_2^{i-1} S_{2,i+1}^n; S_{1,i}|Y_{1,i}S_{2,i}) - n\,\epsilon_n \ , \tag{90}$$

where $(a)$ holds using Fano's inequality (83), $(b)$ holds since the source is memoryless; and $(c)$ holds since $S_{1,i} \ominus (W, Y_1^n, S_2^n, S_{1,i+1}^n) \ominus Y_{2,i+1}^n$ is a Markov chain.

Let $U_{0,i} \triangleq (W, Y_{2,i+1}^n, Y_1^{i-1}, S_2^{i-1}, S_{2,i+1}^n)$ and $U_{1,i} \triangleq Y_{1,i+1}^n$. The above leads to

$$R \geq H(S_2|Y_1) + I(U_0 U_1; S_1|S_2 Y_1) \ , \tag{91}$$

where $U_0 = (U_{0,Q}, Q)$, $U_1 = (U_{1,Q}, Q)$ and $Q \sim \text{Unif}[1:n]$.

Next, we also have

$$nR \geq H(W|Y_2^n) \tag{92}$$

$$\geq I(W; S_1^n S_2^n|Y_2^n) \tag{93}$$

$$= H(S_1^n S_2^n|Y_2^n) - H(S_1^n S_2^n|WY_2^n) \tag{94}$$

$$\overset{(a)}{\geq} H(S_1^n S_2^n|Y_2^n) - H(S_1^n|WS_2^n Y_2^n) - n\epsilon_n \tag{95}$$

$$= nH(S_1 S_2|Y_2) - H(S_1^n|WS_2^n Y_2^n) + H(S_1^n|WS_2^n Y_1^n) - H(S_1^n|WS_2^n Y_1^n) - n\epsilon_n \tag{96}$$

where $(a)$ follows using the Fano's inequality (83).

Let us now define

$$A \triangleq H(S_1^n|WS_2^n Y_1^n) - H(S_1^n|WS_2^n Y_2^n).$$

A key ingredient at this level of the proof is to write the $n$-letter $A$ in single-letter form. To this end, note that we have

$$A = I(S_1^n; Y_2^n|WS_2^n) - I(S_1^n; Y_1^n|WS_2^n) \tag{97}$$

$$= \sum_{i=1}^{n} \left[ I(S_1^n; Y_{2,i}|WY_{2,i+1}^n S_2^n) - I(S_1^n; Y_{1,i}|WY_1^{i-1} S_2^n) \right] \tag{98}$$

$$\overset{(a)}{=} \sum_{i=1}^{n} \left[ I(S_1^n Y_1^{i-1}; Y_{2,i}|WY_{2,i+1}^n S_2^n) - I(S_1^n Y_{2,i+1}^n; Y_{1,i}|WY_1^{i-1} S_2^n) \right] \tag{99}$$

$$\overset{(b)}{=} \sum_{i=1}^{n} \left[ I(S_1^n; Y_{2,i}|WY_{2,i+1}^n Y_1^{i-1} S_2^n) - I(S_1^n; Y_{1,i}|WY_{2,i+1}^n Y_1^{i-1} S_2^n) \right] \tag{100}$$

$$\overset{(c)}{=} \sum_{i=1}^{n} \left[ I(S_{1,i}; Y_{2,i}|WY_{2,i+1}^n Y_1^{i-1} S_2^n) - I(S_{1,i}; Y_{1,i}|WY_{2,i+1}^n Y_1^{i-1} S_2^n) \right] \tag{101}$$

$$= \sum_{i=1}^{n} \left[ H(S_{1,i}|Y_{1,i}WY_{2,i+1}^n Y_1^{i-1} S_2^n) - H(S_{1,i}|Y_{2,i}WY_{2,i+1}^n Y_1^{i-1} S_2^n) \right] \tag{102}$$

$$= \sum_{i=1}^{n} \left[ H(S_{1,i}|Y_{1,i}S_{2,i}U_{0,i}) - H(S_{1,i}|Y_{2,i}S_{2,i}U_{0,i}) \right] \tag{103}$$

where $(a)$ follows using Csiszár-Körner sum-identity

$$\sum_{i=1}^{n} I(Y_1^{i-1}; Y_{2,i}|S_1^n WY_{2,i+1}^n S_2^n) = \sum_{i=1}^{n} I(Y_{2,i+1}^n; Y_{1,i}|S_1^n WY_1^{i-1} S_2^n), \tag{104}$$

($b$) follows using Csiszár-Körner sum-identity

$$\sum_{i=1}^{n} I(Y_1^{i-1}; Y_{2,i}|WY_{2,i+1}^n S_2^n) = \sum_{i=1}^{n} I(Y_{2,i+1}^n; Y_{1,i}|WY_1^{i-1}S_2^n) \ , \tag{105}$$

and ($c$) holds using the fact that, for $j \in \{1,2\}$, the following

$$(S_1^{i-1}, S_{1,i+1}^n) \multimap (S_2^{i-1} S_{2,i+1}^n Y_{2,i+1}^n Y_1^{i-1} W S_{1,i} S_{2,i}) \multimap Y_{j,i} \tag{106}$$

is a Markov chain, the proof of which follows.

The proof of (106) is easy and is as follows. First note that, for $j = 1,2$, we have that

$$(S_1^{i-1}S_{1,i+1}^n S_2^{i-1} S_{2,i+1}^n Y_{2,i+1}^n Y_1^{i-1}) \multimap (S_{1,i}, S_{2,i}) \multimap Y_{j,i} \tag{107}$$

is a Markov chain, which holds since the source is memoryless. Then, given that $W$ is a deterministic function of $(S_1^n, S_2^n)$ one gets that

$$(S_1^{i-1}S_{1,i+1}^n S_2^{i-1} S_{2,i+1}^n Y_{2,i+1}^n Y_1^{i-1} W) \multimap (S_{1,i}, S_{2,i}) \multimap Y_{j,i} \tag{108}$$

is a Markov chain, which in turn implies (106).

Finally, note that we also have

$$H(S_1^n|WS_2^n Y_1^n)$$

$$= \sum_{i=1}^{n} H(S_{1,i}|WS_{2,i}Y_{1,i}Y_1^{i-1}Y_{1,i+1}^n S_2^{i-1}S_{2,i+1}^n S_{1,i+1}^n) \tag{109}$$

$$\overset{(a)}{=} \sum_{i=1}^{n} H(S_{1,i}|WS_{2,i}Y_{1,i}Y_1^{i-1}Y_{1,i+1}^n S_2^{i-1}S_{2,i+1}^n S_{1,i+1}^n Y_{2,i+1}^n) \tag{110}$$

$$\leq \sum_{i=1}^{n} H(S_{1,i}|WS_{2,i}Y_{1,i}Y_1^{i-1}Y_{1,i+1}^n S_2^{i-1}S_{2,i+1}^n Y_{2,i+1}^n) \tag{111}$$

$$= \sum_{i=1}^{n} H(S_{1,i}|S_{2,i}Y_{1,i}U_{0,i}U_{1,i}), \tag{112}$$

where ($a$) follows using the Markov chain

$$Y_{2,i+1}^n \multimap (S_{2,i+1}^n S_{1,i+1}^n Y_{1,i+1}^n S_{2,i}S_2^{i-1}Y_1^{i-1}W) \multimap S_{1,i} \ . \tag{113}$$

which itself holds since

$$Y_{2,i+1}^n \multimap (S_{2,i+1}^n S_{1,i+1}^n Y_{1,i+1}^n) \multimap S_{1,i}S_1^{i-1}S_{2,i}S_2^{i-1}Y_1^{i-1}W \ , $$

is a Markov chain and $W$ is a deterministic function of $(S_1^n S_2^n)$.

Summarizing, combining (96), (103) and (112) we get that

$$R \geq H(S_1 S_2|Y_1) + H(S_1|Y_1 S_2 U_0) - H(S_1|Y_2 S_2 U_0) - H(S_1|Y_1 S_2 U_0 U_1). \tag{114}$$

The end of the proof of converse of Theorem 1 follows by noticing that the so-defined auxiliary random variables $U_{0,i}$ and $U_{1,i}$ satisfy the Markov chain

$$(U_{0,i}, U_{1,i}) \multimap (S_{1,i}, S_{2,i}) \multimap (Y_{1,i}, Y_{2,i}). \tag{115}$$

Also, we have $H(\hat{S}_1|U_0, U_1, Y_1) \leq H(\hat{S}_1|W, Y_1^n) = 0$; and the distortion constraint $\mathbb{E}(d_1(S_1, \hat{S}_1)) \leq D_1$ holds by arguments that are essentially similar to in [6, Appendix A].

### B. Proof of achievability

The proof of achievability of Theorem 1 can be obtained using a coding scheme that is essentially similar to a suitable adaptation of Heegard-Berger coding scheme of [3, Theorem 1, p. 730].

*Codebook generation:*

Set all $S_2^n \in \mathcal{T}_{[S_2]}^{(n)}$ in $2^{nR_2}$ bins $\mathcal{B}^n(w_2)$ indexed by $w_2 \in [1 : 2^{nR_2}]$. For each $S_2^n$, generate $2^{nR_0}$ sequences $u_0^n(w_0)$ each according to $\prod_{i=1}^n P_{U_0|S_2}$, where $w_0 \in [1 : 2^{nR_0}]$, and randomly partition them into $2^{nR'_0}$ bins $\mathcal{B}^n(w'_0)$ that are indexed with $w'_0 \in [1 : 2^{nR'_0}]$. Then, for each $w_0$, generate $2^{nR_1}$ sequences $u_1^n(w_0, w_1)$ each according to $\prod_{i=1}^n P_{U_1|U_0S_2}$ with $w_1 \in [1 : 2^{nR_1}]$, and randomly partition them into $2^{nR'_1}$ bins $\mathcal{B}^n(w'_0, w'_1)$ that are indexed with $w'_1 \in [1 : 2^{nR'_1}]$.

*Encoding:*

In order to encode a pair $(s_1^n, s_2^n)$, the encoder first finds an index $w_2$ such that

$$s_2^n \in \mathcal{B}^n(w_2). \tag{116}$$

It then finds an index $w_0 \in [1 : 2^{nR_0}]$ such that

$$\left(u_0^n(w_0), s_2^n, s_1^n\right) \in \mathcal{T}_{[U_0S_2S_1]}^{(n)}, \tag{117}$$

and an index $w_1 \in [1 : 2^{nR_1}]$ such that

$$\left(u_1^n(w_0, w_1), u_0^n(w_0), s_2^n, s_1^n\right) \in \mathcal{T}_{[U_0U_1S_2S_1]}^{(n)}. \tag{118}$$

It can be shown easily that the error in this encoding step vanishes asymptotically with the block size $n$ as long as

$$R_0 \geq I(U_0; S_1|S_2) \tag{119}$$

$$R_1 \geq I(U_1; S_1|U_0S_2). \tag{120}$$

The encoder transmits the triple $(w_2, w'_0, w'_1)$, i.e, the index of the source component $s_2^n$ as well as the indices of the bins $\mathcal{B}^n(w'_0)$ and $\mathcal{B}^n(w'_0, w'_1)$ in which the found covering codewords $u_0^n$ and $u_1^n$ lie.

*Decoding:*

Decoder 2 reconstructs the sequences $u_0^n$ and $s_2^n$. To this end, it looks for the unique sequences $s_2^n \in \mathcal{B}^n(w_2)$ and $u_0^n(w_0) \in \mathcal{B}^n(w'_0)$ such that

$$\left(u_0^n(w_0), s_2^n, y_2^n\right) \in \mathcal{T}_{[U_0S_2Y_2]}^{(n)}. \tag{121}$$

The error in this step can be made arbitrarily small as long as $n$ is large and

$$R_0 - R_0' \leq I(U_0; Y_2 | S_2) \tag{122}$$

$$R_0 - R_0' - R_2 \leq I(U_0; Y_2 | S_2) - H(S_2 | Y_2). \tag{123}$$

Decoder 1 decodes $s_2^n, u_0^n$ and $u_1^n$. To this end, it looks for a triple $(s_2^n, u_0^n, u_1^n)$ such that $s_2^n \in \mathcal{B}^n(w_2)$, $u_0^n \in \mathcal{B}^n(w_0')$, $u_1^n \in \mathcal{B}^n(w_0', w_1')$ and

$$\left( u_0^n(w_0), u_1^n(w_0, w_1), s_2^n, y_1^n \right) \in \mathcal{T}_{[U_0 U_1 S_2 Y_1]}^{(n)}. \tag{124}$$

Decoder 1 then uses $(u_0^n(w_0), u_1^n(w_0, w_1), y_1^n)$ to find an appropriate estimate $\hat{s}_1^n$ of the sequence $s_1^n$. The error in this decoding step can be made arbitrarily small as long as $n$ is large and

$$R_0 - R_0' \leq I(U_0; Y_1 | S_2) \tag{125}$$

$$R_0 - R_0' - R_2 \leq I(U_0; Y_1 | S_2) - H(S_2 | Y_1) \tag{126}$$

$$R_1 - R_1' \leq I(U_1; Y_1 | U_0 S_2). \tag{127}$$

*Fourier-Motzkin Elimination (FME):*

Finally, applying FME to eliminate $R_0$ and $R_1$ from inequalities $(119) - (127)$, and substituting $R = R_2 + R_0' + R_1'$, yields the desired result

$$R \geq \max\{ I(U_0; S_1 | S_2 Y_1) + H(S_2 | Y_1), I(U_0; S_1 | S_2 Y_2) + H(S_2 | Y_2) \} + I(U_1; S_1 | U_0 S_2 Y_1). \tag{128}$$

# Appendix B
# Proof of Theorem 3

## A. Proof of converse

Let $R$ be a $(D_1, D_2)$-achievable rate for our Heegard-Berger problem with degraded reconstruction sets and common reconstruction of Figure 2. Let $\Phi$ be the associated encoding function, and $g_1$, $g_2$, and $g_s$ the associated reconstruction functions. That is,

$$W = \Phi(S_1^n, S_2^n) \tag{129}$$

$$\hat{S}_1^n = g(W, Y_1^n) \tag{130}$$

$$\hat{S}_{2,1}^n = g_1(W, Y_1^n) \tag{131}$$

$$\hat{S}_{2,2}^n = g_2(W, Y_2^n) \tag{132}$$

with

$$P_e^{(n)} \triangleq \mathbb{P}\left( \hat{S}_{2,1}^n \neq \hat{S}_{2,s}^n \text{ or } \hat{S}_{2,2}^n \neq \hat{S}_{2,s}^n \right) \leq \epsilon_n. \tag{133}$$

First, note that the imposed common encoder-decoders reconstruction constraint for the source component $s_2^n$ implies the following Fano's inequalities,

$$\frac{1}{n} H(\hat{S}_{2,s}^n | \hat{S}_{2,1}^n) \leq \frac{1}{n} + \log_2(\|\hat{S}\|) P_e^{(n)} \quad \text{and} \quad \frac{1}{n} H(\hat{S}_{2,s}^n | \hat{S}_{2,2}^n) \leq \frac{1}{n} + \log_2(\|\hat{S}\|) P_e^{(n)}. \tag{134}$$

Combining (133) and (134), one gets that

$$\frac{1}{n}H(\hat{S}_{2,s}^n|\hat{S}_{2,1}^n) \le \epsilon_n' \quad \text{and} \quad \frac{1}{n}H(\hat{S}_{2,s}^n|\hat{S}_{2,2}^n) \le \epsilon_n'. \tag{135}$$

Next, for the first constraint on the rate $R$, we have

$$nR \ge H(W|Y_1^n) \tag{136}$$

$$\ge I(W; S_1^n S_2^n|Y_1^n) \tag{137}$$

$$= I(W\hat{S}_{2,1}^n; S_1^n S_2^n|Y_1^n) \tag{138}$$

$$= I(W\hat{S}_{2,1}^n; S_1^n S_2^n|Y_1^n) \tag{139}$$

$$\overset{(a)}{\ge} I(W\hat{S}_{2,1}^n\hat{S}_{2,s}^n; S_1^n S_2^n|Y_1^n) - 2n\epsilon_n \tag{140}$$

$$\ge I(W\hat{S}_{2,s}^n; S_1^n S_2^n|Y_1^n) - 2n\epsilon_n \tag{141}$$

$$= \sum_{i=1}^n I(W\hat{S}_{2,s}^n; S_{1,i}S_{2,i}|S_1^{i-1}S_2^{i-1}Y_1^nY_{1,i-1}^n) - 2n\epsilon_n \tag{142}$$

$$= \sum_{i=1}^n I(W\hat{S}_{2,s}^n Y_1^{i-1}Y_{1,i+1}^n S_1^{i-1}S_2^{i-1}; S_{1,i}S_{2,i}|Y_{1,i}) - 2n\epsilon_n \tag{143}$$

$$\overset{(b)}{=} \sum_{i=1}^n I(W\hat{S}_{2,s}^n Y_1^{i-1}Y_{1,i+1}^n S_1^{i-1}S_2^{i-1}Y_2^{i-1}; S_{1,i}S_{2,i}|Y_{1,i}) - 2n\epsilon_n$$

$$\ge \sum_{i=1}^n I(W\hat{S}_{2,s}^{i-1}\hat{S}_{2,s,i+1}^n Y_1^{i-1}Y_{1,i+1}^n Y_2^{i-1}\hat{S}_{2,s,i}; S_{1,i}S_{2,i}|Y_{1,i}) - 2n\epsilon_n \tag{144}$$

where ($a$) holds using (135) and ($b$) holds using the following Markov chain the justification of which will follow,

$$Y_2^{i-1} \;\multimap\!\!\!\!-\; (W, \hat{S}_{2,s}^n, Y_1^{i-1}, Y_{1,i+1}^n, S_1^{i-1}, S_2^{i-1}, Y_{1,i}) \;\multimap\!\!\!\!-\; (S_{1,i}, S_{2,i}). \tag{145}$$

At this stage, we pause to justify (145). We have the following list of Markov chains and implications,

$$(a)\ (Y_2^{i-1}, Y_1^{i-1}, S_1^{i-1}, S_2^{i-1}) \;\multimap\!\!\!\!-\; Y_{1,i} \;\multimap\!\!\!\!-\; (Y_{1,i+1}^n, S_{1,i+1}^n, S_{2,i+1}^n, S_{1,i}, S_{2,i})$$

$$\Rightarrow Y_2^{i-1} \;\multimap\!\!\!\!-\; (Y_{1,i}, Y_1^{i-1}, S_1^{i-1}, S_2^{i-1}) \;\multimap\!\!\!\!-\; (Y_{1,i+1}^n, S_{1,i+1}^n, S_{2,i+1}^n, S_{1,i}, S_{2,i}) \tag{146}$$

$$\overset{(b)}{\Rightarrow} Y_2^{i-1} \;\multimap\!\!\!\!-\; (Y_{1,i}, Y_1^{i-1}, S_1^{i-1}, S_2^{i-1}) \;\multimap\!\!\!\!-\; (W, \hat{S}_{2,s}^n, Y_{1,i+1}^n, S_{1,i+1}^n, S_{2,i+1}^n, S_{1,i}, S_{2,i}) \tag{147}$$

$$\Rightarrow Y_2^{i-1} \;\multimap\!\!\!\!-\; (Y_{1,i}, Y_1^{i-1}, S_1^{i-1}, S_2^{i-1}) \;\multimap\!\!\!\!-\; (W, \hat{S}_{2,s}^n, Y_{1,i+1}^n, S_{1,i}, S_{2,i}) \tag{148}$$

$$\Rightarrow Y_2^{i-1} \;\multimap\!\!\!\!-\; (Y_{1,i}, W, \hat{S}_{2,s}^n, Y_1^{i-1}, Y_{1,i+1}^n, S_1^{i-1}, S_2^{i-1}) \;\multimap\!\!\!\!-\; (S_{1,i}, S_{2,i})\ . \tag{149}$$

where ($a$) holds since the source is memoryless and ($b$) holds since $W$, and so $\hat{S}_{2,s}^n$, are deterministic functions of $(S_1^n, S_2^n)$.

Defining, for $i \in [1:n]$, the auxiliary random variables $U_{0,i} = (W, \hat{S}_{2,s}^{i-1}, \hat{S}_{2,s,i+1}^n, Y_2^{i-1}, Y_{1,i+1}^n)$ and $U_{1,i} = (U_{0,i}Y_1^{i-1})$, the inequality (144) given

$$nR \ge \sum_{i=1}^n I(U_{0,i}U_{1,i}\hat{S}_{2,s,i}; S_{1,i}S_{2,i}|Y_{1,i}) - 2n\epsilon_n. \tag{150}$$

For the second constraint on the rate $R$, we have

$$nR \geq H(W|Y_2^n) \tag{151}$$

$$\geq I(W; S_1^n S_2^n | Y_2^n) \tag{152}$$

$$= I(W\hat{S}_{2,2}^n; S_1^n S_2^n | Y_2^n) \tag{153}$$

$$\geq I(W\hat{S}_{2,2}^n \hat{S}_{2,s}^n; S_1^n S_2^n | Y_2^n) - 2n\epsilon_n \tag{154}$$

$$\geq I(W\hat{S}_{2,s}^n; S_1^n S_2^n | Y_2^n) - 2n\epsilon_n \tag{155}$$

$$= H(S_1^n S_2^n | Y_2^n) - H(S_1^n S_2^n | W\hat{S}_{2,s}^n Y_2^n) - 2n\epsilon_n \tag{156}$$

$$= H(S_1^n S_2^n | Y_2^n) - H(S_1^n S_2^n | W\hat{S}_{2,s}^n Y_2^n) + H(S_1^n S_2^n | W\hat{S}_{2,s}^n Y_1^n) - H(S_1^n S_2^n | W\hat{S}_{2,s}^n Y_1^n) - 2n\epsilon_n \ . \tag{157}$$

The term $[H(S_1^n S_2^n | W\hat{S}_{2,s}^n Y_1^n) - H(S_1^n S_2^n | W\hat{S}_{2,s}^n Y_1^n)]$ on the RHS of (157) can be written as

$$H(S_1^n S_2^n | W\hat{S}_{2,s}^n Y_1^n) - H(S_1^n S_2^n | W\hat{S}_{2,s}^n Y_2^n)$$

$$= I(S_1^n S_2^n; Y_2^n | W\hat{S}_{2,s}^n) - I(S_1^n S_2^n; Y_1^n | W\hat{S}_{2,s}^n) \tag{158}$$

$$\overset{(a)}{=} \sum_{i=1}^{n} \left[ I(S_1^n S_2^n; Y_{2,i} | W\hat{S}_{2,s}^n Y_2^{i-1} Y_{1,i+1}^n) - I(S_1^n S_2^n; Y_{1,i} | W\hat{S}_{2,s}^n Y_2^{i-1} Y_{1,i+1}^n) \right] \tag{159}$$

$$\overset{(b)}{=} \sum_{i=1}^{n} \left[ I(S_{1,i} S_{2,i}; Y_{2,i} | W\hat{S}_{2,s}^n Y_2^{i-1} Y_{1,i+1}^n) - I(S_{1,i} S_{2,i}; Y_{1,i} | W\hat{S}_{2,s}^n Y_2^{i-1} Y_{1,i+1}^n) \right] \tag{160}$$

$$= \sum_{i=1}^{n} \left[ I(S_{1,i} S_{2,i}; Y_{2,i} | U_{0,i} \hat{S}_{2,s,i}) - I(S_{1,i} S_{2,i}; Y_{1,i} | U_{0,i} \hat{S}_{2,s,i}) \right] \tag{161}$$

$$= \sum_{i=1}^{n} \left[ H(S_{1,i} S_{2,i} | U_{0,i} \hat{S}_{2,s,i} Y_{1,i}) - H(S_{1,i} S_{2,i} | U_{0,i} \hat{S}_{2,s,i} Y_{2,i}) \right] \tag{162}$$

where $(a)$ follows using Csiszár-Körner sum identity, applied twice, similar to in (97)–(103), and $(b)$ holds since the following is a Markov chain, the justification of which will follow,

$$(S_1^{i-1}, S_{1,i+1}^n, S_2^{i-1}, S_{2,i+1}^n) \; \multimap \; (U_{0,i}, \hat{S}_{2,s,i}, S_{1,i}, S_{2,i}) \; \multimap \; (Y_{1,i}, Y_{2,i}). \tag{163}$$

We pause to justify (163). This is obtained using the following easy Markov chains and implications,

$$(Y_2^{i-1}, Y_{1,i+1}^n, S_1^{i-1}, S_{1,i+1}^n, S_2^{i-1}, S_{2,i+1}^n) \multimap (S_{1,i}, S_{2,i}) \multimap (Y_{1,i}, Y_{2,i}) \tag{164}$$

$$\overset{(c)}{\Longrightarrow} (W, \hat{S}_{2,s}^n, Y_2^{i-1}, Y_{1,i+1}^n, S_1^{i-1}, S_{1,i+1}^n, S_2^{i-1} S_{2,i+1}^n) \multimap (S_{1,i}, S_{2,i}) \multimap (Y_{1,i}, Y_{2,i}) \tag{165}$$

$$\Longrightarrow (U_{0,i}, \hat{S}_{2,s,i}, S_1^{i-1}, S_{1,i+1}^n, S_2^{i-1}, S_{2,i+1}) \multimap (S_{1,i}, S_{2,i}) \multimap (Y_{1,i}, Y_{2,i}) \tag{166}$$

$$\Longrightarrow (S_1^{i-1}, S_{1,i+1}^n, S_2^{i-1}, S_{2,i+1}) \multimap (U_{0,i}, \hat{S}_{2,s,i}, S_{1,i}, S_{2,i}) \multimap (Y_{1,i}, Y_{2,i}) \tag{167}$$

and where $(c)$ follows since $W$, and so $\hat{S}_{2,s}^n$, are deterministic functions of $(S_1^n, S_2^n)$.

Finally, we terminate the proof of converse of Theorem 3 by noticing that the reconstruction $\hat{S}_{1,i} = g(W, Y_1^n)$ clearly satisfies $\hat{S}_{1,i} = g'(U_{0,i}, U_{1,i}, Y_{1,i})$ for some function $g'$. $\qquad \square$

## B. Proof of achievability

The proof of achievability of Theorem 3 is as follows. First, we show the achievability of the following rate distortion function,

$$R(D_1, D_2) = \min \max\{I(U; S_1 S_2 | Y_1), \ I(U; S_1 S_2 | Y_2)\} + I(U_1; S_1 S_2 | Y_1 U) \tag{168}$$

where the minimization is over all conditionals $P_{UU_1 | S_1 S_2}$ satisfying that $(U, U_1) \multimap (S_1, S_2) \multimap (Y_1, Y_2)$ is a Markov chain and there exist functions $\phi$ and $\psi$ such that:

$$\mathbb{E} d_1(S_1, \hat{S}_1) \leq D_1 \tag{169}$$

$$\mathbb{E} d_2(S_2, \hat{S}_2) \leq D_2 \tag{170}$$

with

$$\hat{S}_2 = \psi(U) \tag{171}$$

$$\hat{S}_1 = \phi(Y_1, U, U_1). \tag{172}$$

Next, we evaluate the above region with the choice $U = (U_0, \hat{S}_2)$ to recover the result of Theorem 3.

*Codebook generation:*

Generate $2^{nR_0}$ sequences $u^n(w_0)$ following $\prod_{i=1}^n P_U$ where $w_0 \in [1 : 2^{nR_0}]$ and set them in $2^{nR'_0}$ bins $\mathcal{B}^n(w'_0)$ with $w'_0 \in [1 : 2^{nR'_0}]$. Then, for each $w_0$, generate $2^{nR_1}$ sequences $u_1^n(w_0, w_1)$ following $\prod_{i=1}^n P_{U_1 | U}$ with $w_1 \in [1 : 2^{nR_1}]$ and set them in $2^{nR'_1}$ bins $\mathcal{B}^n(w'_0, w'_1)$ with $w'_1 \in [1 : 2^{nR'_1}]$. Also, fix a reconstruction function $\psi$ such that: $\mathbb{E} d_2(S_2, \psi(U)) \leq D_2$ and denote $\psi^n$ its n-letter extension such that $\hat{s}_{2,s}^n = \psi^n(u^n(w_0))$.

*Encoding:*

Upon observing $S_1^n$ and $S_2^n$, find an index $w_0 \in [1 : 2^{nR_0}]$ such that:

$$\left(u^n(w_0), s_2^n, s_1^n\right) \in \mathcal{T}_{[US_2S_1]}^{(n)} , \tag{173}$$

and an index $w_1 \in [1 : 2^{nR_1}]$ such that:

$$\left(u_1^n(w_0, w_1), u^n(w_0), s_2^n, s_1^n\right) \in \mathcal{T}_{[UU_1S_2S_1]}^{(n)} . \tag{174}$$

The encoder transmits $w'_0$ and $w'_1$,i.e., the indices of the bins in which $u^n$ and $u_1^n$ lie. This encoding step has small error as long as $n$ is large and

$$R_0 \geq I(U; S_1 S_2) , \tag{175}$$

$$R_1 \geq I(U_1; S_1 S_2 | U) \tag{176}$$

*Decoding:*

Decoder 2 reconstructs the sequences $\hat{s}_2^n$. To this end, it first looks for the unique sequence $u^n(w_0) \in \mathcal{B}^n(w_0')$ such that

$$\left( u^n(w_0), y_2^n \right) \in \mathcal{T}_{[UY_2]}^{(n)} . \tag{177}$$

Then, it sets $\hat{s}_{2,2}^n = \psi^n(u^n(w_0))$ as the final reconstruction. (Note here that the reconstruction $\hat{S}_2$ can not depend on the available side information sequence).

The error in this decoding step can be made arbitrarily small as long as $n$ is large and

$$R_0 - R_0' \le I(U; Y_2) , \tag{178}$$

Decoder 1 looks for a couple of sequences $u^n \in \mathcal{B}^n(w_0')$ and $u_1^n \in \mathcal{B}^n(w_0', w_1')$ verifying:

$$\left( u^n(w_0), u_1^n(w_0, w_1), y_1^n \right) \in \mathcal{T}_{[UU_1Y_1]}^{(n)} , \tag{179}$$

Then, from $(u^n(w_0), u_1^n(w_0, w_1), y_1^n)$ the decoder can recover the reconstruction sequences $\hat{s}_{2,1}^n = \psi^n(u^n(w_0))$ and $\hat{s}_1^n = \phi(u^n(w_0), u_1^n(w_0, w_1), y_1^n)$.

Similarly, the error in this decoding step can be made arbitrarily small as long as $n$ is large and

$$R_0 - R_0' \le I(U; Y_1) , \tag{180}$$

$$R_1 - R_1' \le I(U_1; Y_1 | U) . \tag{181}$$

Note that, in this scheme, if the encoding and decoding steps are performed correctly are successful, then all terminals can reconstruct $\hat{s}_2^n$ with essentially no error.

The rest of the proof follows by a standard application of FME to eliminate $R_0$, $R_1$ and $R_2$ from inequalities (175) – (181) and substituting $R = R_0' + R_1'$ to get the desired result. $\qquad\square$

## References

[1] A. D. Wyner and J. Ziv, "The rate-distortion function for source coding with side information at the decoder," *Information Theory, IEEE Transactions on*, vol. 22, pp. 1–10, 1976.

[2] A. Sgarro, "Source coding with side information at several decoders," *Information Theory, IEEE Transactions on*, vol. 23, no. 2, pp. 179–182, 1977.

[3] C. Heegard and T. Berger, "Rate distortion when side information may be absent," *Information Theory, IEEE Transactions on*, vol. 31, no. 6, pp. 727–734, 1985.

[4] R. Timo, T. Oechtering, and M. Wigger, "Source coding problems with conditionally less noisy side information," *Information Theory, IEEE Transactions on*, vol. 60, no. 9, pp. 5516–5532, 2014.

[5] A. Kimura and T. Uyematsu, "Multiterminal source coding with complementary delivery," *arXiv preprint arXiv:0804.1602*, 2008.

[6] R. Timo, A. Grant, and G. Kramer, "Rate-distortion functions for source coding with complementary side information," in *Information Theory Proceedings (ISIT), 2011 IEEE International Symposium on*. IEEE, 2011, pp. 2934–2938.

[7] S. Watanabe, "The rate-distortion function for product of two sources with side-information at decoders," *Information Theory, IEEE Transactions on*, vol. 59, no. 9, pp. 5678–5691, 2013.

[8] R. Timo, T. Chan, and A. Grant, "Rate distortion with side-information at many decoders," *Information Theory, IEEE Transactions on*, vol. 57, no. 8, pp. 5240–5257, 2011.

[9] C. Tian and S. N. Diggavi, "On multistage successive refinement for wynerziv source coding with degraded side informations," *Information Theory, IEEE Transactions on*, vol. 53, no. 8, pp. 2946–2960, 2007.

[10] ——, "Side-information scalable source coding," *Information Theory, IEEE Transactions on*, vol. 54, no. 12, pp. 5591–5608, 2008.

[11] Y. Steinberg, "Coding and common reconstruction," *Information Theory, IEEE Transactions on*, vol. 55, no. 11, pp. 4995–5010, 2009.

[12] D. Slepian and J. K. Wolf, "Noiseless coding of correlated information sources," *IEEE Trans. Inf. Theory*, vol. 19, pp. 471–480, 1973.

[13] T. Berger, *Rate Distortion Theory: A Mathematical Basis for Data Compression*. Prentice-Hall, 1971. [Online]. Available: https://books.google.co.uk/books?id=-HV1QgAACAAJ

[14] T. Weissman and A. El Gamal, "Source coding with limited-look-ahead side information at the decoder," *Information Theory, IEEE Transactions on*, vol. 52, no. 12, pp. 5218–5239, 2006.

[15] B. N. Vellambi and R. Timo, "The Heegard-Berger problem with common receiver reconstructions," in *Information Theory Workshop (ITW), 2013 IEEE*. IEEE, 2013, pp. 1–5.

[16] B. Ahmadi, R. Tandon, O. Simeone, and H. V. Poor, "Heegard-Berger and cascade source coding problems with common reconstruction constraints," *Information Theory, IEEE Transactions on*, vol. 59, no. 3, pp. 1458–1474, 2013.